\documentstyle[preprint,aps]{revtex}
\begin{document}
\draft
\preprint{TIFR/TH/93-42}
\title
{Supercooling and Nucleation in Phase Transitions \\
 of the Early Universe}
\author{B. Banerjee and R. V. Gavai}
\address
{Tata Institute of Fundamental Research, Homi Bhabha Road,
Bombay 400 005, India}
\date{\today}
\maketitle
\begin{abstract}
The three phase transitions -- the GUT, the electro-weak and the
quark-hadron, which the universe is assumed to have undergone
produce very important physical effects if they are assumed
to be of first order. It is also important that enough supercooling
is produced at these transitions so that the rate of nucleation
of the lower temperature phase out of the higher temperature phase
is large. We argue on the basis of finite-size scaling theory
that for the quark-hadron and the electro-weak phase transitions
the universe does not supercool enough to give sizeable nucleation
rates. Only  for the GUT transition the nucleation probability
seems to be significant.
\end{abstract}

\narrowtext

According to the standard big bang model, the universe started from
 a state of very high density and temperature.
Due to the expansion of the
universe the temperature falls and depending on the underlying theory
of particle interactions, a sequence of phase transitions takes place.
Physical consequences of the three such phase transitions have been
extensively investigated.  These are 1) the GUT phase transition
at a~$T_{c1} \sim O(10^{14}-10^{16)}$~GeV, where the symmetry
between the
strong and the electroweak interactions is spontaneously broken,
 2) the
electroweak (EW) phase transition at a~$T_{c2}\sim O(10^2)$~GeV,
 where the
electroweak symmetry is spontaneously broken, and  finally
 3) the chiral
symmetry breaking and/or the confining QCD phase transition at
a~$T_{c3} \sim O(10^{-1})$~GeV. The value of $T_{c1}$
is, of course, purely conjectural as there is no viable grand
unified theory
at the moment.  However, non-perturbative results from simulations of
corresponding lattice field theories indicate that the values of
the other
two critical temperatures,~$T_{c2}$~ and~$T_{c3}$~are reasonably well
determined.

It is natural to expect some cosmological and astrophysical
consequences
of these phase transitions. Indeed, the GUT phase transition
leads to the
formation of various topological structures-- domain walls, cosmic
strings, magnetic monopoles etc.\cite{wei}.  It has been also
exploited in the
inflationary scenarios for the early universe. Recently there
have been
attempts to show that the baryon asymmetry of the universe
can be generated
at the electroweak phase transition \cite{coh,farr}. There have
been also
speculations about the creation of initial density inhomogeneities
necessary for large scale structure formation in the universe
at this transition \cite{dol}. The
quark-hadron transition has been shown \cite{appl,full} to lead
to an alternative
scenario for nucleosynthesis with substantially large
baryon density contrast, $\Omega_B$.
In many
of these applications, a (strong) first order phase transition with
substantial supercooling has been assumed.  The consequent bubble
nucleation is then the  mechanism responsible for the
expected effects.

For the purposes of this article we will therefore assume
that all the phase
transitions to be of first order and examine critically the
estimates of
the corresponding nucleation rates. It is usually assumed that
the high
temperature symmetric phase, A, goes into a metastable state and is
`supercooled' before decaying into droplets of the less symmetric low
temperature phase, B.  The universe passes through this series of
equilibrium and metastable states only if the interactions necessary
for
particle  distribution functions  to adjust to the changing
temperature
are rapid compared to the expansion rate of the universe.  A rough
criterion that a reaction rate is fast enough for maintaining
equilibrium
is~$\Gamma > H$~where ~$\Gamma$~is the interaction rate per
particle and
$H$ is the Hubble constant. If~$\Gamma < H$~then the particles
 `freeze
out' and do not contribute to the maintenance of equilibrium
\cite{wei}.

The rate of nucleation of droplets of phase B has so far been
calculated
using the homogeneous nucleation theory \cite{land}. It is assumed
that the droplets
of phase B arise through spontaneous thermodynamic fluctuations
in phase
A. For the formation of a spherical droplet of radius $r$ the
change in the free energy of the system is given by,
\begin{equation}
\label{ass}
 \Delta F = {4\pi\over 3} {\left(p_A (T)-p_B(T)\right) r^3}
 + 4\pi r^2 \sigma \end{equation}
where $p_A(T)-p_B(T)$ is the difference in pressure in the two phases
at temperature $T$ and $\sigma$ is the surface tension of the
interface of
the phases. $\Delta F$ increases with $r$ till a maximum value
$r_{cr}$ is reached where
\begin{equation}
\label{bill}
r_{cr}={2\sigma\over{p_B(T)-p_A(T)}}
\end{equation}
Droplets with $r<r_{cr}$ shrink and disappear while droplets with
$r>r_{cr}$ grow. The rate of nucleation per unit volume is given by
\begin{equation}
\label{cat}
I= I_0 exp \left(-{\Delta F_{cr}\over T} \right)
\end{equation}
where $\Delta F_{cr}$ is the value of $\Delta F$ for
$r=r_{cr}$ and $I_0$
is the prefactor. For small supercooling, that is, for
{}~$\eta=(T_c-T)/T_c<$ 1
Eq.\ (\ref{cat}) can be written as
\begin{equation}
\label{dog}
I=I_0 exp\left(-{16\pi\over 3} \label{w}
{\sigma^3\over{T_c \Delta E^2 \eta^2}}
\right)
\end{equation}
where $\Delta E$ is the latent heat per unit volume.  Thus the rate
decreases exponentially, as supercooling becomes smaller, unless
$I_0$ happens to be very large.

Since the early days of the nucleation theory
much effort has gone into the calculation of $I_0$ \cite{fren}.
The most
well-known result is that due to Becker and D\"oring \cite{beck}.
They calculated $I_0$ by considering explicitly the kinetics of the
condensation process.  In recent years Langer \cite{lang} has
developed a more
detailed theory of nucleation based on statistical mechanical
considerations. He obtains an expression for $I_0$ very different
from
that of Becker and D\"oring. However the numerical values of $I_0$
do not differ much in the two theories \cite{turs}.

For the quark-hadron phase transition, Fuller et al.\cite{full}
calculated
the rate by setting $I_0=T_c^4$.
Recently Csernai and Kapusta \cite{kapu} have calculated $I_0$
using Langer's theory. They obtain
\begin{equation}
\label{eel}
I_0 = {16\over3\pi}{\left(\sigma\over3T\right)^{3/2}}
{\sigma\eta_A
r_{cr}\over{\xi_A^4(\Delta w)^2}} \end{equation}
where $\eta_A$ and $\xi_A$ are respectively the shear viscosity and
a correlation length in the phase A and $\Delta w$ is the
difference in
the enthalpy densities in the two phases. It turns out
that
the new values of $I_0$ are {\it smaller} than the prefactor
$T_c^4$ for
the quark-hadron phase transition \cite{kapu}.  Therefore, the
crucial parameter which
governs the nucleation rate in this case is indeed the possible
amount of
supercoooling the universe can undergo near $T_c$. It seems
likely that these arguments apply to other, especially
the elecroweak, phase transitions as well.

Before we address the question of the possible amount of
supercooling the
universe can undergo at any of these transitions, it is perhaps
worthwhile
to point out that even in a simple heterogeneous nucleation
mechanism, it
is still the supercooling which dominates the nucleation rate.
As an example of heterogeneous nucleation we consider the
condensation
of light quarks on heavy quarks in the case  of quark--hadron
transition
in analogy with the condensation of water vapour on ions \cite{flet}.
The contribution to the change free energy of the system is now
given by
\begin{equation}
\label{fox}
\Delta F_{\rm impurity} = \alpha \bigg( 1 - {\varepsilon_A \over
\varepsilon_B} \bigg)\bigg({1 \over r} - {1 \over r_0} \bigg)
\end{equation}
where $\varepsilon_A$ and $\varepsilon_B$ are the dielectric constants
in the A and B phases
, $r_0$ is the effective radius of the impurity particles, in this
case the heavy quarks and $\alpha$ is the QCD coupling constant.
The  Coulomb-like form of the potential in Eq. (\ref{fox}) is justified
for a deconfined quark-gluon plasma at sufficiently high
temperatures.
Minimizing
the full $\Delta F$, one can obtain the corresponding $r_{cr}$
and the
modified nucleation rate.  For small $\alpha$, one
obtains
\begin{equation}
\label{goat}
r^{het}_{cr} = r_{cr}\left(1-
{\alpha\over{8\pi \sigma r^3_{cr}}}\right).
\end{equation}
 From the above equation, one sees that the critical radius for the
 heterogeneous system is smaller than the
$r_{cr}$  for the homogeneous case, but it still is governed
by $\eta$, the magnitude
of supercooling, in the same way as before.  Thus in the limit of
vanishing
supercooling the critical radius of the stable bubble is still
too large
to allow significant nucleation rates. Of course, for very small
$\eta$ the solution Eq. (\ref{goat}) is not valid but then it is easy to
show that the full $\Delta F$ has no minimum at all.

This leads us to the central question which we wish to discuss
in this paper: is it possible to
estimate the amount of supercooling the universe can undergo at a
phase
transition ?  We suggest that standard arguments from the finite size
scaling theory \cite{barb,bind}  near a first order phase transition
can be exploited to
answer this question. If $\xi(T)$ is the correlation length in a
given
phase at a temperature T close to the transition point and $L$ is the
linear size of the system then clearly for $L \le \xi$ one expects
large
finite size effects to cause rounding of a discontinuity and
broadening of
the transition region whereas for $L \gg \xi$, the system should
behave as
if it is in the thermodynamic limit. Indeed, while the system
could remain
trapped in a single, perhaps metastable, phase for the former case,
it
will be in a mixed phase of several domains of both phases for the
latter.
In fact, Challa et al \cite{chal} in their study of temperature-driven
first order transitions have shown  that
\begin{equation}
\label{hare}
{\Delta T \over T_c} \simeq { T_c \over \Delta E L^3 }.
\end{equation}
This result can be expressed in terms of $\xi$ and $\Delta E$,
by observing that
$\Delta E$, the latent heat per unit volume can be written as
\begin{equation}
\label{ibex}
\Delta E = { c_1 T_c \over { \xi}^3}, \end{equation}
purely on dimensional grounds. Here $c_1$ is a  constant.
( Note that the above equation is consistent with the statement that
a second order transition is the limit of a first order
phase transition for infinite correlation length  or vanishing latent
heat.)
Substituting Eq. (\ref{ibex}) in Eq. (\ref{hare})  we get,
\begin{eqnarray}
\label{jack}
 \eta &=& A{ \bigg({ \xi \over L} \bigg)^3} \nonumber \\
      &= & A{ \bigg({ \xi T_c  \over  L T_c} \bigg)^3},
\end{eqnarray}
where the constant of proportionality, $A$, should be typically
 $O(1)$.
This relation has been verified for the quenched QCD and both
$A$ and
$\xi T_c$ have been estimated \cite{gav} to be $O(1)$.  For the
electro-weak theory
or any grand unified theory, no such test has so far been made and no
estimate of $\xi T_c$ or $A$ is available. On the other hand Eq. (9)
has been verified for a large number of models in statistical
 mechanics \cite{bind}. It thus seems natural to assume that for
 electro-weak and GUT
theories also both these dimensionless numbers are $O(1)$, although
we will allow them to vary up to $O(10^2)$.

Thus  for determining the amount of supercooling at a phase transition
in the
universe one needs the correlation length and the volume
of the universe
at the crtical temperature. The volume can be obtained from standard
cosmology. Assuming an ideal gas equation of
state for the matter in the early universe,
\begin{equation}
\label{koil}
\rho = 3 P = {\pi^2 \over 30} N(T) T^4~~,~~\end{equation}
where $N(T) = N_B(T) + 7/8 N_F(T)$ is the total number of bosonic (B)
and fermionic (F) degrees of freedom. The age of the universe at
temperature
$T$ is given by the relation \cite{wei}
\begin{equation}
\label{lamb}
 t = {1 \over 4 \pi} \sqrt{{45 \over \pi N(T)}}
{ M_P \over T^2}~~.
\end{equation}
Here $M_P \sim 10^{19}$ GeV is the Planck mass.  Since the radius
of the
universe as given by the particle horizon is $3t$, the volume, V,
of the
universe at  temperature T is given by
\begin{eqnarray}
\label{moos}
V T^3 &=& {405 \over 16} \sqrt{ 45\over \pi^7 N^3(T)}
\bigg( {M_P \over T}
\bigg)^3  \nonumber \\
&=& {5.625 \times 10^{57}} \over {N^{3/2}(T) T^3}
\end{eqnarray}
Here T is expressed in GeV.
Substituting Eq. (\ref{moos}) in Eq. (\ref{jack}), one finds that
the supercooling which the universe can undergo is
\begin{equation}
\label{nix}
\eta = 1.78 \times 10^{-58} A (\xi T_c)^3 N^{3/2}(T_c) T
        ^3 ~~.
\end{equation}

As mentioned earlier, all quantities in Eq. (\ref{nix}) are known only
for the
quark-hadron phase transition.  Using the data from lattice QCD
for $T_c$
and $\xi(T_c)T_c$ \cite{gav} and $N(T_c)=51.25$ [corresponding
to photons(2),
gluons(16), electrons(4), muons(4), neutrinos(6) and two flavours of
quarks(24)], one  finds that the supercooling $\eta$ is negligibly
small, as shown in our earlier work \cite{ban} where all the caveats
in using the
lattice QCD data and their effects are also discussed.
This result is not changed if we take the new results
of Ref. 12 for $I_0$ or even the possibility of heterogeneous
nucleation caused by heavy quarks.

Similar arguments as given above can be applied to {\it any} other
phase
transition occurring during the evolution of the universe. Thus,
for the
electroweak phase transition, $N(T_c)$  is 51.5
[corresponding to
W and Z-bosons(6), taus(4), and four flavours of quarks(48)]
and we obtain $T_c \sim 250$ GeV on simple dimensional grounds.
 One-loop perturbation theory
yields \cite{ande} $\xi(T_c)T_c =28.1$ and $T_c = 184 $ GeV for a Higgs
mass of 80 GeV, while lattice investigations \cite{hell} of the
$O(4)$ model suggest $T_c =370$ GeV for a Higgs mass close to its
triviality bound of about 650 GeV.
Choosing the estimates of Ref. 20, we find that
\begin{equation}
\label{owl}
\eta_{\rm EW} = 1.57 \times 10^{-43} A ~~,\end{equation}
which can be significant only for unnaturally large A.
 Note that the
expected uncertainty of a factor of two in $T_c$ and a correlation
length
which is 2-3 orders of magnitude larger than the one used
above will not alter
the conclusion at all.  Furthermore, the ideal gas equation of state
used here should be adequate, since the non-perturbative
contributions to the
energy density are unlikely to change $N(T_c)$ by more than an
order of magnitude.

Finally, for the GUT phase transition, one sees that substituting
$T_c=T_{c1}$, the small numerical factor in Eq. (\ref{nix}) is increased
by 42-48 orders of magnitudes. Since $N(T)$ may increase by a further
factor of 2 or so, the universe can undergo significant supercooling
at
this phase transition provided the correlation length is
$\sim 100T_c$. Indeed it seems that  only for theories which have
either a transition temperature close to $M_P$ or a very large
(but finite) correlation length near $T_c$ it is possible to get
sufficient
supercooling and hence interesting effects through bubble nucleation.

\acknowledgments

B. Banerjee thanks D. N. Schramm for the hospitality at the
Astronomy and Astrophysics Center, University of Chicago,
 where a part of this work was done. His visit was supported
by NSF grants INT 91-101171DS and INT 87-15411.

\end{document}